\documentclass[vecphys]{svmult}

\usepackage{graphicx}        
\usepackage[bottom]{footmisc}

\newcommand{\msun}{M$_\odot$}
\newcommand{\rsun}{R$_\odot$}
\newcommand{\kms}{km~s$^{-1}$}

\begin{document}

\title*{The Bizarre Spectral Variability of \\ Central Stars of Planetary Nebulae}
\author{Orsola De Marco\inst{1},
S. Wortel\inst{1}, Howard E. Bond\inst{2}, and Dianne Harmer\inst{3}}
\authorrunning{De~Marco et al.} 
\institute{American Museum of Natural History \texttt{orsola@amnh.org}
\and Space Telescope Science Institute \texttt{bond@stsci.edu}
\and National Optical Astronomical Observatories \texttt{diharmer@noao.edu}}
%
%
\maketitle

\begin{abstract}
A radial velocity (RV) survey to detect central stars in binary systems was carried out between 2002 and 2004. \cite{DeMarco2004} reported that 10 out of 11 monitored stars exhibited strong RV variability, but periods were not detected. Since other mechanisms, such as wind variability, can cause apparent RV variations, we monitored 4 of the 10 RV-variable stars at echelle resolutions to determine the origin of the variability. Although RV changes are confirmed for all four stars, none of them can be ascribed to binarity at this time. However, only for IC~4593 is wind variability able to explain most (though not all) spectral variability. For BD+33~2642, no wind and no pulsations appear to be the origin of the RV changes. Finally, M~1-77 and M~2-54, both known to be irregular photometric variables, exhibit dramatic RV and line shape variability of the hydrogen and He~I absorption lines, as well as large RV variability of weaker lines, which do not change in shape. There is no satisfactory explanation of this variability, though a combination of wind variability and pulsations is still the best guess at what makes these stars so variable. We suggest that luminous central stars are ill suited to detect spectroscopic binaries, because winds (and possibly pulsations) are pervasive and would mask even strong periodicities. It it likely that a sample of intrinsically faint central stars would more readily yield binary information.

\end{abstract}

\section{Introduction}
\label{sec:introduction}
Thirteen out of $\sim$100 central stars of planetary nebulae (CSPNe) were found to be binaries (\cite{Bond2000}) with periods smaller than 3 days (though most have periods smaller than only one day). These are all post-common envelope (CE) binaries discovered via periodic photometric variability caused by irradiation effects, ellipsoidal variability or because eclipses occur. The predicted post-CE binary period distribution changes depending on the assumed value of the CE efficiency parameter (e.g., \cite{Yungelson1993}, \cite{Han1995}). Form these predictions, we expect  $\sim$25-65\% of all CSPNe to be in binaries with periods smaller than 3 months, that can be detected by RV surveys.  

A RV survey carried out by \cite{DeMarco2004} reported 10 out of 11 monitored CSPNe to be RV variables. For none periods could be found, but the cadence of the monitoring was not ideal. Since wind variability could be the cause of RV shifts we initiated a monitoring campaign at echelle resolutions to scrutinize line shape changes which could reveal the origin of the RV shifts.  


\section{Observations, Reductions and Analysis}
\label{sec:obsredana}

Our Kitt Peak 4-m telescope 2005, May 22-27, and 2006, July 18-24, echelle spectroscopy afforded us a resolution of 4.5~\kms\ over the  wavelength range 3900--5575~\AA. The weather and seeing were very poor. The signal-to-noise ratio (SNR) of the single exposures varied between 20 and 40, sufficient for BD+33~2642, because of the numerous lines in this spectrum. For the other three stars, with fewer and/or broader absorption and emission lines, we averaged three exposures achieving a SNR in the range 40-70. 
    
To determine relative RV shifts we used the cross-correlation technique implemented by the IRAF task {\it fxcor}, which Fourier transforms two  continuum-subtracted spectra, after applying some user-specified filtering. In some cases we chose specific wavelength ranges where lines of interest resided. This is useful when different lines exhibit different behaviors and need individual testing. Relative RV shifts are converted to absolute ones (shifts with respect to the laboratory line center) by measuring the wavelength of the lines of interest in the template spectrum using Gaussian fits, correcting for the heliocentric systemic velocity (measured from the PN lines) and heliocentric correction.  

From the PN lines we can assess that the wavelength calibration is accurate to $\sim$0.5~\kms. To this we add in quadrature the uncertainty from fitting the cross-correlation function. The latter is determined using a Monte Carlo test similar to that adopted by \cite{Armandroff1995}. Details will be given in De~Marco et al. (in preparation).

\section{Results}
\label{sec:results} 

\begin{figure}
\vspace{1.5in}
\centering
\includegraphics{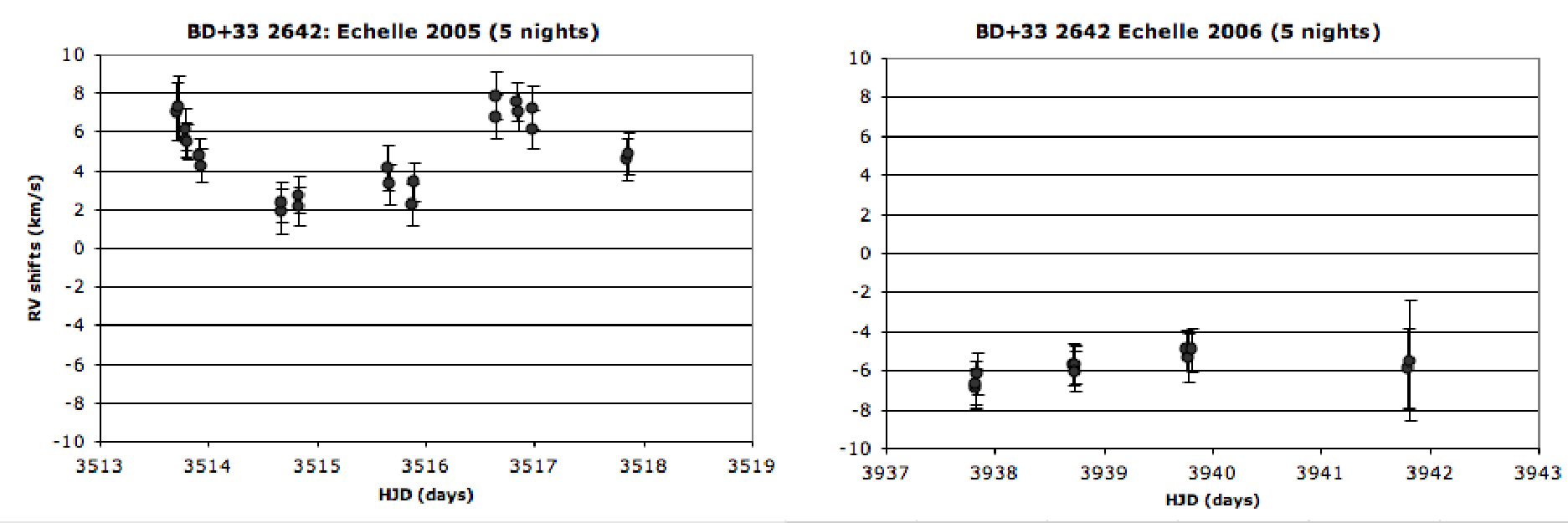}
\caption{Absolute (with respect of the lines' laboratory wavelengths) RV shifts for the absorption line spectrum of BD+33~2642. The heliocentric systemic velocity was measured from one PN line (H$\beta$) to be --93~\kms.}
\label{fig:bd+33}
\end{figure}

{\bf BD+33~2642} is a well known photometric standard (\cite{Landolt2007}). \cite{Napiwotzki1994} discovered it to be the central star of a 5'' halo PN, and determined $T_{\rm eff}$=20\,200~K, $\log g$=2.9, a spectroscopic mass of 0.560~\msun\ and $R = 4.40$~\rsun.  \cite{Stalio1980} analyzed its UV spectrum and determined this object to have a weak wind, with a mass-loss rate $<$4.7$\times$10$^{-8}$~\msun~yr$^{-1}$, large for this type of star.  

The echelle observations confirm the RV behavior reported by \cite{DeMarco2004}.  In 2005 (Fig.~\ref{fig:bd+33}) we see a clear trend with a seemingly periodic fluctuation with a 6~\kms\ amplitude, but 
centered at $\sim$5~\kms\ instead of zero, as would be expected for a binary motion. In 2006 this trend is not obvious and all data points are offset by about --10~\kms\ with respect to the 2005 data (making the mean RV shifts close to zero - Table~\ref{tab:results}). Fitting the entire data-set does not return any believable period. A fit to the 2005 data returns a period of $\sim$4~days, but with low confidence.  

The absorption lines show no line shape variability, which could be ascribed to wind variability and could cause the RV shifts. Pulsations are not likely to cause the variability since this star is not photometrically variable above $\sim$0.01~mag \cite{Landolt2007}. It is at this stage impossible to determine what causes the variability in this star.  

{\bf IC~4593} is an O5f(H) star. It was analyzed by \cite{Mendez1990} who found it to have $T_{\rm eff}$ = (40\,000 $\pm$ 4000)~K and $\log g$ = (3.5$\pm$0.2). \cite{Patriarchi1995} found it to be a wind variable from UV line profile changes. The 12''-PN has two ansae (jets) protruding in opposite directions and suggesting that if this star were a binary, the line of sight would be near  the equatorial plane.  

The spectra are extremely variable. We measured RV shifts by selecting out three line sets: (i) the three He~II absorption lines $\lambda\lambda$4200,4542,5412, (ii) the He~II emission line at 4686~\AA, or (iii) the spectral range 4620--4680~\AA, which contains C~III-IV and N~III emission lines. The first two line sets show large RV amplitude variation (50 and 60~\kms, respectively), while the third set, likely to be the least affected by wind variability, show very little variation (amplitude about 5~\kms) although the trend of this variation is similar to that of the He~II line at 4686~\AA. From this latter line set we deduce that this star is not a binary to within the precision of the observations (unless, of course, the orbital plane is close to the plane of the sky - in contrast to what one might imagine looking at the PN morphology). 

It is odd that {\it all} shifts of the He~II absorption lines are red of the lines' centers, since, typically, absorption lines affected by winds are shifted to the blue. It is also strange that at some epochs, the He~II emission line is blue-shifted, again not expected from wind theory.  In conclusion, while 
the line changes in IC~4593 appear to be due to wind variability, winds do not readily explain all of the changes. Pulsations might have something to do with some of the changes, but, as we will discuss later, the relationship between pulsations and RV changes is not a straight forward one.

%
%

{\bf M~1-77 and M~2-54} have similar spectra. They are cool central stars ($T_{\rm eff} \sim$25\,000~K) of low-excitation (no [O~III] lines are present) small and young PNe (\cite{Sabbadin1983}, \cite{Acker1992}). They are photometric variables. M~2-54 was found to have two multi-hour periods. \cite{Handler1995} and \cite{Handler2003} ascribe these variability to pulsations and claim the discovery of a new class of pulsators, which they call the  ZZ Lep  stars. \cite{Maene1994}, however, call these stars irregular variables, not being convinced that the variability, which changes over months to year time-scales, is due to pulsations. Winds are also not thought to be the cause of the irregular photometric variability,  since for one star in this class (IC~418), IUE spectral variability did not correlate with photometric variability. The IUE spectra of these stars show no emission lines, indicating that the winds are not strong.  

We only took echelle observations of M~1-77 and M~2-54 in 2006, resulting in 5 usable spectra per star with SNR=40-65, each created from the average of three exposures. The spectra  display Balmer emission lines due to the PN, as well as Balmer absorption lines due to the star. We also observe a wealth of absorption lines due primarily to He~I and O~II, as well as a handful of weak emission lines, such as C~II $\lambda$4267. Since the hydrogen and He~I line shapes are extremely variable, we measured the RV shifts by selecting wavelength ranges containing only O~II and other weak metal lines, whose shapes remained approximately constant in time.

\begin{figure}
\vspace{2.0in}
\centering
\includegraphics{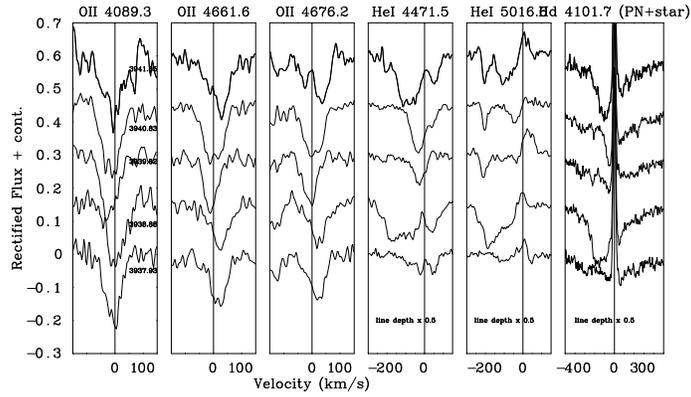}
\caption{A selection of lines from the spectra of M~1-77 taken on 5 subsequent nights.}
\label{fig:m177spectra}        
\end{figure}

\begin{figure}
\vspace{1.5in}
\centering
\includegraphics{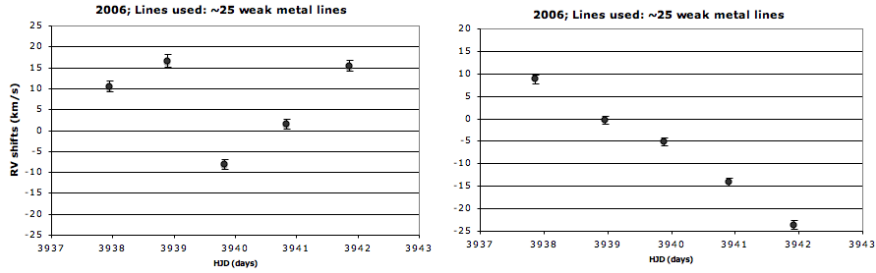}
\caption{Absolute (with respect of the lines' laboratory wavelengths) RV shifts for the weak metal absorption lines of M~1-77 (left) and M~2-54 (right). From the PN lines of these two CSPN we measured heliocentric systemic velocities of --68.4 and --67.7~\kms, respectively.}
\label{fig:m177-m254}       
\end{figure}

In Fig.~\ref{fig:m177spectra} we show a selection of lines and their changes over the 5-night observing period. Hydrogen and He I absorption lines vary extremely between a day and the next, with their absorption profiles splitting and changing in strength. There is almost no variability on a time-scale of about one hour, judging from the individual exposures. Weaker metal lines do not appear to vary in shape to within the SNR of the observations and we use them for our RV measurements. In Fig.~\ref{fig:m177-m254} we show that the absolute RV shifts of the weak metal lines have quite large amplitudes. These shifts are different from those that we would obtain if we measured the He~I lines.  Binarity, wind variability or pulsations do not readily explain the RV behavior of the weak metal lines, although it is probable that the interplay of pulsations and variable mass-loss might, in the end, explain these spectral changes. 

\begin{table}
\centering
\caption{Variability diagnostics for the central stars.}
\label{tab:results}
\begin{tabular}{lccccl}
\hline\noalign{\smallskip}
Star & Mean RV & $\sigma_{RV}$ & Mean Error & \# data  & Lines Used\\
       &  (\kms)     &  (\kms)           & (\kms)        &     points            & \\
\noalign{\smallskip}\hline\noalign{\smallskip} 
IC~4593        
                     & 3.8 & 13.5 & 4.0   &14 &He~II abs. $\lambda$$\lambda$4200,4541,5412\\
                     & 11.4 & 18.1 & 1.6   &14 &He~II em. $\lambda$4686\\
                     & 1.5  & 3.0 & 0.95   &14 &C~III-IV, N~III em. $\lambda$4630-4665\\
BD+33$^a$ 
                    & 1.5   & 5.4   & 1.2   &34   & Entire Spectrum\\
M~1-77         
                    & 7.2  & 10.4 & 1.3   &5  & Weak Metal Lines\\ 
M~2-54         
                    &--6.8  & 12.5  &0.93  &5   & Weak Metal Lines\\
\noalign{\smallskip}\hline
\end{tabular}
$^a$BD+33~2642

\end{table}




\section{Conclusion}
\label{sec:conclusion}
We have no conclusive results as to the binary status of these 4 central stars. BD+33~2642 is a real mystery, with very little, if any, photometric variability, yet a non-periodic RV amplitude of $\sim$10~\kms. There is also no line shape change which might betray a wind origin for the variability. IC~4593 is the best characterized case: there is a very strong RV variability likely due to winds. Some of the emission lines least affected by winds show almost no variability, confirming that this star is likely not a spectroscopic binary. There remain however doubts as to why some of the absorption lines are shifted to the red of center and why some of the emission lines are shifted blue of center (both behaviors are contrary to wind theory).

M~1-77 and M~2-54 are almost identical stars. They are thought to be pulsators by some (\cite{Handler2003}), but not by others (\cite{Maene1994}). So the peculiar and extreme He~I and hydrogen line variability might or might not be related to the photometric variations and might or might not be due to pulsations. The RV variations of the O~II lines, which do not vary in shape as do the hydrogen and He~I lines, and whose RV shifts are not correlated with those of the hydrogen and He~I lines, are even more difficult to understand. The RV shifts of the weak metal lines are not periodic, but have a distinct trend. In conclusion,  these stars' spectral changes while not easy to interpret, could be due to the interplay of pulsations and winds. Binarity, however, cannot yet be excluded for these two stars.

It is likely that wind and pulsational variability would be small or non-existent for hotter and dimmer CSPNe. The presence of winds is a strong function of the stellar luminosity and pulsations in the hotter CSPNe is limited to the hydrogen-deficient atmospheres of PG1159 stars (\cite{Corsico2006}). A sample of hot and dim, hydrogen-normal CSPNe  would make better targets for RV surveys aimed at detecting the central star spectroscopic binary fraction. Unfortunately, these would also be apparently faint, and such a survey would necessitate 8-m class telescopes.



\end{document}